\documentstyle[12pt,epsfig,amssymb]{article}

\begin{document}
\setlength{\parskip}{0.45cm}
\setlength{\baselineskip}{0.75cm}
\begin{titlepage}

\vspace{1cm}
%
\vspace{0.5cm}
\begin{center}
\Large

{\bf A common origin of all the species of \\
high-energy cosmic rays?}\\
\vspace{0.5cm}
\large
 Arnon Dar$^{a,b}$, A. De R\'ujula$^a$ and Nikos Antoniou$^{a,c}$\\

\vspace*{0.5cm}
\normalsize
$^a$ Theory Division, CERN, CH-1211 Geneva 23, Switzerland \\
$^b$ Technion, Israel Institute of Technology,
Haifa 32000, Israel\\
$^c$ Department of Physics, University of Athens,
15771 Athens, Greece \\
\vspace{0.7cm}
%
\large
{\bf Abstract} \\
\end{center}
\vspace*{-0.5cm}
\noindent

Cosmic ray nuclei, cosmic ray electrons
with energy above a few GeV, and the
diffuse gamma-ray background radiation (GBR) above a few MeV,
presumed to be extragalactic, could all have their
origin or residence in our galaxy {\it and its
halo}. The mechanism accelerating hadrons and electrons
is the same, the electron spectrum is modulated by inverse
Compton scattering on starlight and on the
microwave background radiation;  the $\gamma$-rays are the resulting
recoiling photons.  The spectral indices of the cosmic-ray
electrons and of the GBR, calculated on this simple
basis, agree with observations. The angular dependence
and the approximate magnitude of the GBR are also explained.

\vspace{1.5cm}


\vspace{1.5cm}

\end{titlepage}
\newpage
\normalsize
%
The origin and spectrum of non-solar cosmic ray hadrons, electrons
and photons has
been debated for almost a century. The
cosmic ray (CR) nuclei have a power-law
spectral flux ${\rm dF/dE\propto E^{-\beta_i}}$
with a series of break-point
energies ($\beta_{1,2,3}$ $\sim 2.7,~3.0$, and 2.5 in the intervals
${\rm  10^{10}~eV < E < E_{knee}}$ $\sim 3 \times 10^{15}$ eV,
$~{\rm E_{knee} < E < E_{ankle}}$ $\sim 3 \times 10^{18}$ eV,
and
${\rm E_{ankle} < E < E_{GZK}}$ $\sim 6\times 10^{19}$ eV).
Below ${\rm E_{knee}}$, protons constitute  $\sim 96\%$
of the nucleon flux at fixed energy per nucleon. Their flux is
\cite{BG}:
 \begin{equation}
{\rm {dF_p\over dE_p}\simeq 1.8 \left
[E_p \over GeV \right]^{-2.70\pm 0.05}
~cm^{-2}~s^{-1}~sr^{-1}~GeV^{-1}}.
\label{protons}
\end{equation}
A presumed
extragalactic origin~\cite{Protheroe} of the highest-energy
CRs is at variance
with observations \cite{hecr} of CRs above the `GZK cutoff': $\gamma$-rays 
and neutrinos are not good candidate primaries~\cite{H?} and,
for ${\rm E > E_{GZK}}$, the flux of protons and nuclei
originating more than $\sim\! 20$ Mpc away \cite{Lee}
should be quenched~\cite{GZK} by
the cosmic background radiation (CBR).
In this letter we argue that,
above typical nuclear energies of a few MeV, non-solar
CR electrons and the  `diffuse'
$\gamma$-ray background may be intimately related to hadronic CRs,
whose observed spectral shape we use as input.

By assuming a common accelerating mechanism for
all charged particles we deduce a power law
that fits, above ${\rm E_e=5}$ GeV, the CR electron
flux~\cite{eflux0}~\cite{eflux1}
shown in Fig.~1, from the hadron spectral index $\beta_1$. Unlike protons,
electrons are significantly affected by
Compton scattering on starlight and on the CBR.
The  up-scattered photons have a spectral index that
agrees with that of the observed~\cite{Sree} diffuse $\gamma$-ray
background  above a few MeV, shown in Fig. 2.
To obtain the observed $\gamma$-ray intensity, we need a
scale-height of the CR electrons much bigger
than the accepted $\sim 1$ kpc
minimum~\cite{Broad}.
There is evidence for CR electrons well above
galactic disks~\cite{Duric} from  observations of synchrotron
emission by galaxies seen edge-on. We envisage the possibility
that the `box' in which charged CRs reside be much larger than
the galactic disk, and we call this box `the halo'~\cite{DP}.

A classic argument determines the volume
of a region to which CRs are
confined. An abundance ratio of secondary to primary rays (e.g.,
Li/C) is used to deduce ${\rm X}$: the mean column density  they  
traverse.  Ratios of  unstable to stable isotopes
(e.g. $^{10}$Be/$^{9}$Be or $^{26}$Al/$^{27}$Al) are used
to deduce the mean length, ${\rm L}$, of their voyage.
 The inferred mean density~\cite{Simp} is
$\bar \rho= \rm{X/L\sim \rho(disk})/5$. The usual
conclusion --that the confining volume is $\sim 5\, {\rm V(disk})$--
relies on tacit assumptions that need not be right.
If the halo is magnetized, as the disk is, we have a
``leaky box'' within a (much less dense)
leaky box. If the lifetimes of $^{10}$Be
and $^{26}$Al are not longer than the mean
residence time in the disk, the observed ratios
determine ${\rm L(disk})$, but not ${\rm L(halo})$.
The stable isotope ratios do
determine the total column density ${ \rm X(tot})$, but this may
include various prolonged journeys in the halo. Without
knowing ${\rm L(halo})$ the density $\bar \rho= {\rm X(tot)/L(tot})$
of the overall confining domain cannot be estimated.
The conventional tacit assumption
is that the stable CRs that reach us cannot have spent long periods
in a baryonically thin halo, before or between disk-crossings.

To relate the spectra of CR electrons and protons
we shall need an estimate of the protons' spectrum at their source.
At ${\rm E<E_{knee}}$ a source spectrum
${\rm dF^s/dE}$ with index ${\rm\beta_s\sim 2.2}$
is obtained from collisionless-shock simulations~\cite{BO}
or analytical estimates of acceleration by relativistic jets~\cite{D98}.
The CR spectrum is modulated by their
residence time in the galaxy, ${\rm \tau_{gal}(E)}$. For a steady source of
CRs the energy dependence of the observed flux is roughly that of
${\rm \tau_{gal}\, dF^s/dE}$.
Observations of astrophysical and solar plasmas and of nuclear abundances 
as functions of energy
indicate~\cite{Swordy} that ${\rm \tau_{gal}(E)\propto E^{ -0.5\pm 0.1}}$,
explaining $\beta_1\sim 2.7$, as in Eq.~(\ref{protons}).

The CR electron flux of Fig.~1
is well fit, from  $\sim 10$ GeV to $\sim 2$ TeV,  by:
\begin{equation}
{\rm {dF_e\over dE_e}\simeq (2.5\pm 0.5)\times 10^5 \left
[E_e \over MeV \right]^{-3.2\pm 0.10}
~cm^{-2}~s^{-1}~sr^{-1}~MeV^{-1}}.
\label{electrons}
\end{equation}
The $e^+$ admixture is $\sim 7\%$
from 3 to 50 GeV~\cite{Barb}. To a good
approximation, as we assume at all energies,
CR electrons are not pair produced, nor are they
secondary products of interactions, nor
the result of relic-particle annihilation or decay.

Practically all CR acceleration mechanisms invoke a region of space
that is swept by a moving magnetic field, such as would
be carried by the rarefied plasma in a supernova shell~\cite{Protheroe}
or by a `plasmoid' of jetted ejecta~\cite{DP}.
The magnetic field acts as a moving `mirror' that imparts the
same distribution in velocity, or Lorentz factor
${\rm \gamma=E/m\,c^2}$, to all charged
particles. To the extent that particle-specific losses
(such as synchrotron radiation) can be neglected
at the acceleration stage, all source fluxes
have the same energy-dependence. Confinement effects preserve
this equality for ultrarelativistic electrons and protons: their
behaviour in a magnetic maze is the same.
But, unlike for hadrons, the `cooling' time of electrons in their
`inverse' Compton scattering (ICS)
interactions with starlight and the CBR is shorter than their galactic
confinement time,
${\rm \tau_{gal}(E)}$, above a relatively low energy.

Consider electron interactions with the CBR.
Let ${\rm T_0=2.728}$
K, ${\rm n_0\approx 411~cm^{-3}}$, and
${\rm \epsilon_0\approx 2.7\, kT_0\approx 6.36\times 10^{-10}~ MeV}$
be the current CBR temperature, number density and mean  
energy~\cite{MF}. The cross section for ICS
is ${\rm \sigma_{_T}\approx 0.65\times 10^{-24}~cm^2}$
(for the relevant ${\rm E_e}$ range the Thompson limit is accurate,
even for ICS on starlight).
The mean energy ${\rm E_\gamma}$ of the upscattered photons,
--or ${\rm \Delta E_e}$, the mean
energy loss per collision-- is
${\rm E_\gamma \approx \Delta E_e \approx (4/3)\epsilon_0
(E_e/ m_ec^2)^2}$.
The single electron interaction rate is
${\rm \sigma_{_T}\, n_0\, c}$.
Let R (also an inverse time) be the
production rate of CR electrons, assumed to
be constant~\cite{BG}, and let ${\rm dn^s_e/dE_e}$ be their source
number-density spectrum.
The actual density ${\rm dn_e/dE_e}$
in an interval ${\rm dE_e}$ about ${\rm E_e}$ is
continuously replenished and depleted
by electrons whose energy is being degraded by ICS. This leads to
a steady-state situation in which production and losses are in balance:
\begin{equation}
{\rm
\sigma_{_T}\, n_0\, c\, {d\over dE_e} \left(E_\gamma\, {dn_e\over  
dE_e}\right)=
      R\,{dn^s_e\over dE}}\, .
\label{balance}
\end{equation}

For a relatively uniform galactic CR
electron density, Eq.~(\ref{balance}) also applies to the
local electron flux ${\rm dF_e\simeq (c/4\pi)dn_e}$.
Substitute the spectrum ${\rm
dn^s_e/dE\sim E_e^{-\beta_s}}$
into the flux version of Eq.~(\ref{balance}) to
obtain:
\begin{equation}
{\rm {dF_e\over dE_e}= {3\, m_e^2\, c^4\, R\over
4\,(\beta_s-1)\,c\,\sigma_T\, n_0\,\epsilon_0\,
 E_e}\; { dF^s_e\over dE_e}\propto E_e^{-(\beta_s+1)}}\, .
\label{electron}
\end{equation}
Compton scattering on starlight and synchrotron radiation on
magnetic fields can be included by replacing $n_0\,\epsilon_0$
by the total energy density $\Sigma n_i\epsilon_i+B^2/(8\pi)$,
whose local variations do not affect the spectral shape.
For electrons with ${\rm E_e<(m_e/m_p)\, E_{knee}}$$\sim 1.6$ TeV
we deduced that
$\beta_s\sim 2.2~.$ Thus, $\beta_s+1=3.2$,
in agreement with the observed spectral index,
Eq.~(\ref{electrons}) and Fig.~1.
Above the `electron's knee' at ${\rm E_e\sim 1.6}$ TeV
the spectrum should steepen up by $\Delta
\beta \simeq 0.25$, like that of CR hadrons~\cite{D98}. The available 
spectral measurements extend only to ${\rm E_e\leq 1.5}$ TeV.

The cooling time of electrons in the CBR bath is given by
\begin{equation}
{\rm \tau_{_{ICS}}(E_e)\simeq {E_e\over
n_0\,\sigma_{_T}\, c\, \Delta E_e}\simeq {3\,m_e^2\, c^4\over
 4 \, E_e\, n_0\,\epsilon_0\,
\sigma_{_T}}\simeq 1.2\times \left [{E_e\over GeV}\right]^{-1}~Gy}\, .
\label{coolingbis}
\end{equation}
At our location in the Galaxy, starlight and magnetic
fields have energy densities similar to that of the CBR,
and ${\rm \tau_{_{ICS}}}$ is shorter by a factor $\sim 3$.
The galactic escape time of GeV electrons should be
similar to that of CR protons
${\rm \tau_{gal}(E)\propto E^{ -0.5\pm 0.1}}$ \cite{Swordy},
and has a weaker energy dependence than $\rm \tau_{_{ICS}}$ does.
At sufficiently low energy, then, ${\rm \tau_{gal}<\tau_{ICS}}$,
and processes other than ICS cooling become relevant.
 The slope of the electron spectrum of Eq.~(\ref{electron})
should change as the energy is lowered. The observed spectrum of
Fig.~1 shows such a change, but it occurs below ${\rm E_e=10}$ GeV,
a range in which local modulations would mask the effect.

Data from the SAS 2 satellite~\cite{TF} suggested the existence
of an isotropic, `diffuse' gamma background radiation (GBR).
The EGRET instrument on the
Compton Gamma Ray Observatory
has confirmed its existence.
By removal of the Galactic-disk diffuse emission
a uniformly distributed GBR has been found, of alleged
extragalactic origin~\cite{Sree}.
The GBR flux in the observed energy range of ${\rm 30~MeV}$ to ${\rm  
120~GeV}$
is well described by a single power law:
\begin{equation}
{\rm {dF_\gamma\over dE_\gamma}\simeq (2.74\pm 0.11)\times 10^{-3} \left
[E_\gamma \over MeV \right]^{-2.10\pm 0.03}
~cm^{-2}~s^{-1}~sr^{-1}~MeV^{-1}}\; .
\label{photons}
\end{equation}
Many possible sources for the GBR have been discussed, and are reviewed
in~\cite{Sree}.

EGRET data on $\gamma$-rays above 1 GeV show an excess over the expectation
from cosmic-ray production of $\pi^0$s~\cite{PE98}.
Electron bremsstrahlung in gas is not the source
of the 1--30 MeV inner-Galaxy $\gamma$-rays observed by
COMPTEL~\cite{Strong}, since their latitude distribution is
broader than that of the gas.
These findings imply  that a source such as ICS may be
more important than previously believed~\cite{Moska}.
We go even further and propose that the GBR is
not mainly extragalactic, but dominantly
due to the ICS of CBR and of starlight
photons by CR electrons in the galactic halo. It is useful to discuss 
the CBR and starlight contributions in turn.

Let us characterize the CR electron density in the Galaxy and its halo
as a smooth spherical
distribution. Assume the electron
flux, Eq.(\ref{electrons}), observed at ${\rm E_e>10}$ GeV,
to be representative of the local interstellar value.
Define a `scale radius' for the CR
electrons,
${\rm R_e \equiv \int dr [dn_e(r)/dr]/n_e(0)}$.
Let $\theta$ be the angle away from the direction to the
galactic center, distant ${\rm d_\odot\sim 8.5}$ kpc.
The CR-electron density distribution in
energy and position may be
convoluted with the  CBR spectrum to derive the
resulting flux of ICS up-scattered photons~\cite{FMRL}.
For ${\rm R_e^2 \gg d_\odot^2}$ the result is insensitive to the spatial
shape of the electron distribution and
very well approximated by:
\begin{equation}
{\rm {dF_\gamma\over dE_\gamma}\simeq {n_0~\sigma_{_T}~R_e\;
f(\theta)\over 2~E_\gamma}
\left[ E_e~ {dF_e\over dE_e}\right]_{E_e=
\overline{E}_e};~~~
\overline{E}_e\equiv m_ec^2
\sqrt{3\, E_\gamma\over4\,\epsilon_0}}\; ,
\label{ICSphotons}
\end{equation}
where ${\rm \overline{E}_e}$ is obtained by inverting
${\rm E_\gamma}$. The form-factor $f(\theta)$
 reflects our excentricity:
\begin{equation}
{\rm R_e\;f(\theta)}={\rm
d_\odot\cos\theta+\sqrt{R_e^2-d_\odot^2\sin^2\theta}}
\; .
\label{asymmetry1}
\end{equation}
For the value we shall estimate, ${\rm R_e=30}$ kpc, ${\rm  
f(\theta)=1\pm 0.28}$, for $\theta=0,\pi$.

Substitute Eq.~(\ref{electrons}) into Eq.~(\ref{ICSphotons}) to obtain:
\begin{equation}
{\rm {dF_\gamma^{CBR}\over dE_\gamma}
\simeq (1.43\pm 0.30)\times\! 10^{-3}
\left[{R_e\, f(\theta)\over 30\, kpc}\right] \left
[E_\gamma \over MeV \right]^{-2.10\pm 0.05}
cm^{-2}s^{-1}sr^{-1}MeV^{-1}}.
\label{ICSphotons2}
\end{equation}
The index coincides with the measured one
and, for  ${\rm R_e=30~kpc}$, the
normalization is roughly half of what is observed, Eq.(\ref{photons}).
The electron spectrum of Eq.~(\ref{electrons})
describes the data in the range ${\rm E_e>5}$ GeV, so that
Eq.~(\ref{ICSphotons2}) should be valid above ${\rm E_\gamma\sim
100}$ keV. We have predicted a steepening of the electron spectrum
at ${\rm E_e\sim 1.6}$
TeV. By the same token, the spectral index $\beta\sim 2.1$
of Eq.~(\ref{ICSphotons2}) should increase by a currently unobservable
$\Delta \beta \sim 1/8$ at ${\rm E_\gamma\sim 10}$ GeV.

It is difficult to model in detail ICS on
starlight~\cite{Hun}~\cite{Sree}. We make a coarse estimate of the effect,
at high latitudes, of an electron distribution with large $\rm{R_e}$.
Approximate the Galaxy's starlight, of average energy $\epsilon_\star\sim
1$ eV, as that produced by a source at its center with the galactic
luminosity ${\rm L_*}=2.4\times 10^{10}$ ${\rm L_\odot}$. Assume that
${\rm R^2_e>>d^2_\odot}$ and replace in Eq.~(\ref{ICSphotons2})  the halo column density of CBR photons in the $\theta$ direction, 
${\rm n_0 R_e f(\theta)}$, by the same quantity for starlight: 
\begin{equation}
{\rm N_\star\approx {L_*\over 4 \pi c \epsilon_\star}\int_0^{\infty} {dx\over
d_\odot^2 -2xd_\odot cos\theta+x^2}= {L_*\over 4 \pi c \epsilon_\star}
\left[{(\pi\!-\!\theta) \over d_\odot\sin\theta}\right]},
\label{SLcolumndensity} 
\end{equation} 
to obtain 
\begin{equation} 
{\rm {dF_\gamma^{SL}\over dE_\gamma} \simeq
(1.14\pm 0.24)\times\! 10^{-3} \left[{2(\pi\!-\!\theta) \over
\pi\sin\theta}\right] \left [E_\gamma \over MeV \right]^{-2.10\pm 0.05}
cm^{-2}s^{-1}sr^{-1}MeV^{-1}}. 
\label{ICSphotons3} 
\end{equation} 
The point source approximation is not accurate enough for directions close
to the galactic disk and galactic bulge, but the GBR data~\cite{Sree} mask
$\rm{|b|\le 10^o}$ as well as $\rm{|b|\le 30^o}$ for $\rm{|l|\le 40^o}$. 

The angular average of the sum of CBR and starlight contributions 
to the GBR,
Eqs.(\ref{ICSphotons2}) and (\ref{ICSphotons3}), is shown in Fig.~2.
It has the observed spectral shape and, for $R_e=30$ kpc,
the observed magnitude. The fitted value $\rm{R_e}\sim 30$ kpc
is imprecise: the starlight to CBR ratio
is proportional to $\epsilon_\star/\epsilon_0$ raised to a
poorly determined power, $0.10\pm 0.05$.
Our spectral index is
independent of direction, as observed, see Figs.~5 and 6 of ~\cite{Sree}. 
The spectral-index data are given in~\cite{Sree} for 36 $\rm{(b,l)}$ domains, 
in Figs.~3-7 we have redrawn them as functions of $\theta$: 9 values for
each half hemisphere.
The statistical test for a flat distribution is surprisingly good:
${\rm \chi^2/d.o.f.\sim 0.5}$.

The intensity of the GBR
diminishes as the angle away from the galactic center
increases, see Figs.~5 and 6a of~\cite{Sree}, which we
have redrawn as functions of $\theta$ in Figs.~8-12.
Even after eliminating the four data points closest to
the galactic center (at $\bar\theta=49.2^o$), about which the
observers feel unsure~\cite{Sree},  ${\rm \chi^2/d.o.f.=1.8}$ for
constant intensity.  The sum of Eqs.~(\ref{ICSphotons2}) and (\ref{ICSphotons3}) fits better the observed angular trend of the GBR, 
as shown in Figs.~8-12 (${\rm \chi^2/d.o.f.=1.3}$). This agreement would  
presumably further improve with a more realistic treatment of starlight. 

For $\rm{R_e}= 30$ kpc, the luminosity of our galaxy in
$\gamma$-rays of energy above E is:
\begin{equation}
{\rm L_E \simeq (1.5\pm 0.5)\times 10^{40}
\left[ {E\over MeV}\right]^{-0.10\pm 0.05}~erg~s^{-1}}.
\end{equation}
This result is slowly converging and uncertain, but
a future $\gamma$-ray telescope, such as GLAST, could possibly
see the corresponding glow of Andromeda's halo.

To estimate the contribution of external galaxies to the
GBR, some concepts and numbers need be recalled.
Hubble's `constant' is ${\rm H_0=100~h~km~s^{-1}Mpc^{-1}}$,
with ${\rm h\sim 0.65}$;
${\rm \Omega_m}$ and $\Omega_\Lambda$ are matter and vacuum
cosmic densities in critical units;
${\rm \Omega\equiv\Omega_m+ \Omega_\Lambda}$;
$y\equiv 1+z$ is the redshift factor.
In a Friedman model, the time to redshift relation is
${\rm dy/dt=-H_0\, f(y)\, y}$, with
${\rm f(y)\equiv [(1-\Omega)\, y^2+\Omega_m\, y^3+\Omega_\Lambda]^{1/2}}$.
The luminosity density of the local universe~\cite{Love} is
${\rm \rho_{_L}\simeq 1.8\times 10^8\, h\, L_\odot\, Mpc^{-3}}$. The
combination ${\rm \rho_{_L}/L_*}$ provides an estimate of the average
number density of `Milky-Way-equivalent' galaxies. If the sources
of CRs are young supernova remnants or gamma-ray bursts,
the CR production rate ought to be proportional~\cite{Wijers}
to the star formation rate ${\rm R_{SFR}}$,
recently measured up to redshift
${\rm z\simeq 4.5}$~\cite{S}.

The energy of CBR photons up-scattered by electrons at `epoch y'
is proportional
to ${\rm T(y)=y~T_0}$ and is subsequently redshifted by the same factor;
hence the corresponding spectra from distant galaxies have
the same energy dependence as from our galaxy
 up to ${\rm E_\gamma\sim 10}$ GeV, where
absorption on the infrared background becomes relevant~\cite{SS}.
The cosmological starlight contribution is
redshifted and consequently becomes negligible.
For the sum of all galaxies, we obtain:
\begin{equation}
{\rm {dF_\gamma\over dE_\gamma}\simeq
{dF^{CBR}_{\gamma}\over dE_\gamma}\,\left [1+
{4\,\pi\, R_e^2\over3}~{\rho_{_L}\over L_*}~{c\over H_{0}}
\int_1 {R_{SFR}(y)\over R_{SFR}(0)}~{y\over f(y)}~{dy\over y^3} \right]},
\label{allgals}
\end{equation}
where the CBR flux is that of Eq.~(\ref{ICSphotons2}).
For ${\rm R_{SFR}}$ we interpolate the summary
values of \cite{S}.
For ${\rm \Omega=\Omega_m=1}$ and ${\rm R_e=30}$ kpc, the
extragalactic
contribution in Eq.~(\ref{allgals})  is at the 4\% level (the
h-dependences cancel). Other sensible values of  the
cosmological parameters give similar results.

We have given a simple explanation for the observed spectral index of CR
electrons. We have suggested that these electrons populate a ``halo''
region well beyond the disk
of our Galaxy and that they produce the GBR by
inverse Compton scattering on CBR and starlight photons. Thus,
the diffuse $\gamma$ radiation (in directions other than those
of the Galaxy's disk and bulge) 
would not originate outside the Galaxy, but outside its starry domain.
We have demonstrated that the predicted spectral index
and angular distribution of the produced radiation agree with the GBR
data, their normalization results in an estimate of the scale-height of
the electron distribution.
Our results can be checked in various ways.  The power index
$\beta$ of the electron spectrum should steepen above ${\rm E_e\approx
1.6}$ TeV by $\Delta\beta\sim 1/4$.  The halo of Andromeda should shine in
gamma rays above a few MeV. The GBR should reflect the asymmetry of our
off-center position in the galactic disk and its energy spectrum should
not have the GZK-like cutoff expected~\cite{SS} for cosmological sources.


\newpage
\begin{figure}
\begin{center}
\vspace*{-1.6cm}
\hspace*{-1cm}
\epsfig{file=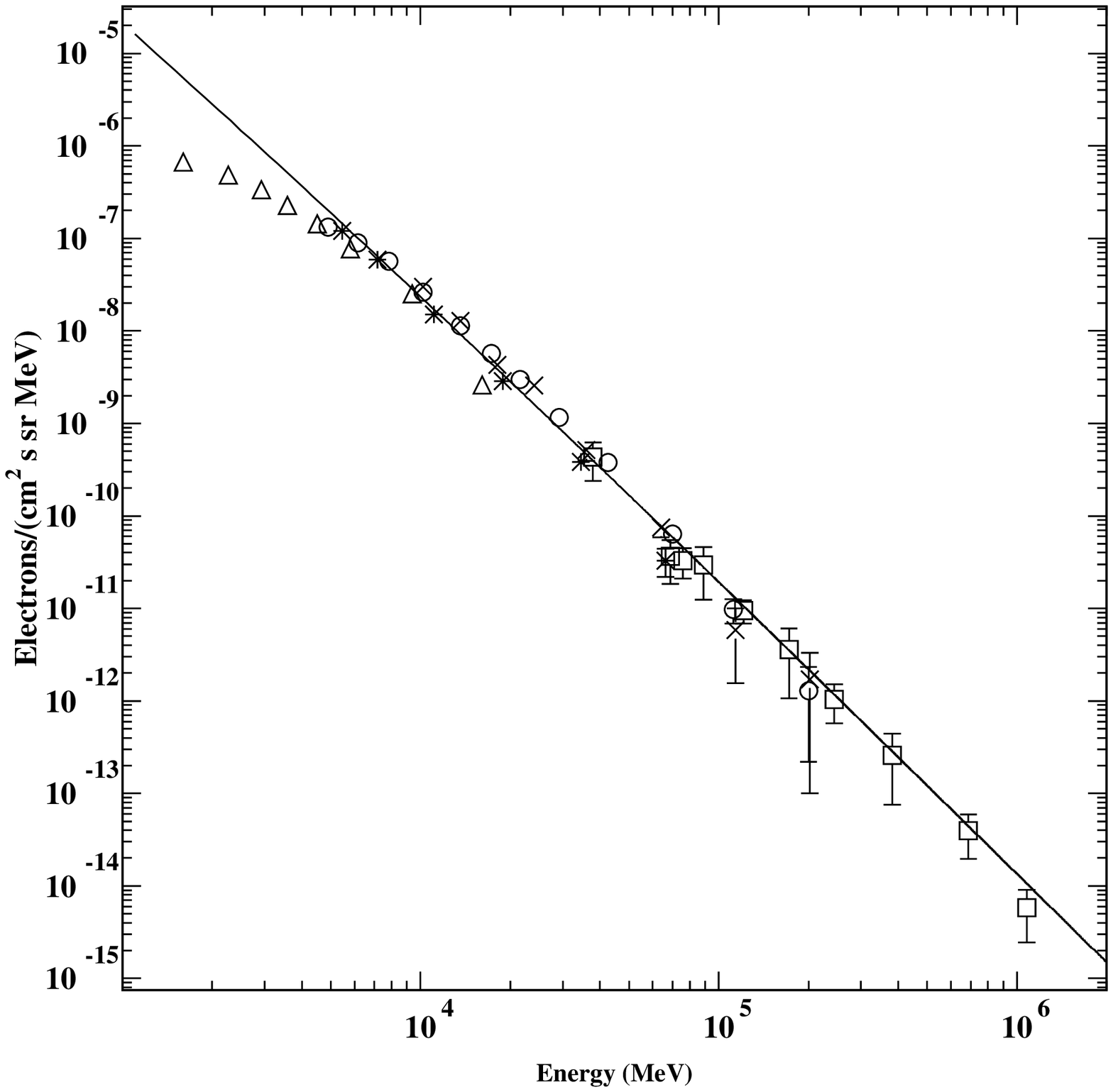,width=15cm}
\caption{The primary cosmic ray electron spectrum as
measured~\cite{eflux1} by
Prince [crosses], Nishimura {\it et al.} [squares], Tang  [circles],
Golden {\it et al.} [triangles] and  Barwick {\it et al.}
[stars]. The theoretical curve is normalized to the data.}
\vspace*{-0.5cm}
\end{center}
\end{figure}

\newpage
\begin{figure}
\begin{center}
\vspace*{-1.6cm}
\hspace*{-1cm}
\epsfig{file=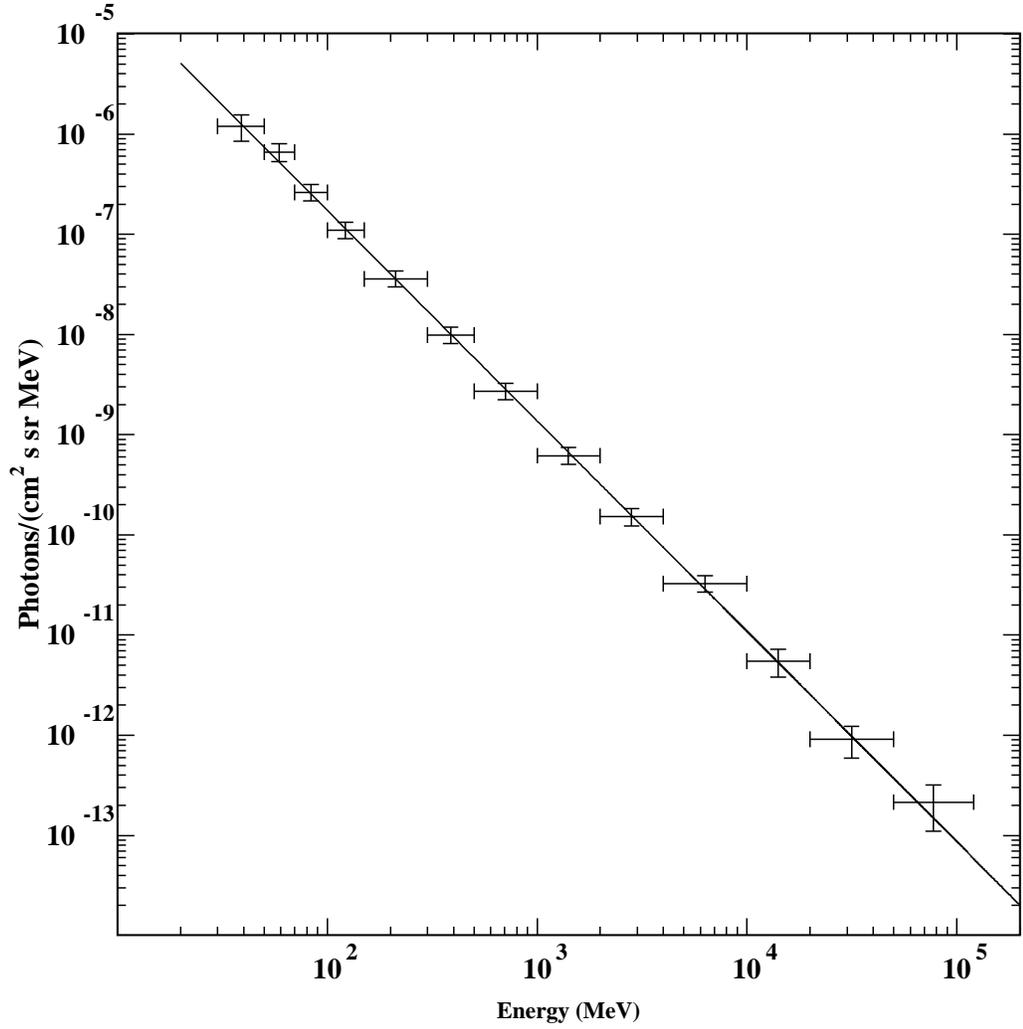,width=15cm}
\caption{Comparison between the spectrum  of
the GBR, measured by EGRET~\cite{Sree},
and the prediction for ICS of starlight and the CMB by CR
electrons. The slope is the central prediction, the normalization
is the one obtained for ${\rm R_e}= 30$ kpc.}
\vspace*{-0.5cm}
\end{center}
\end{figure}

\newpage
\begin{figure}
\begin{center}
\vspace*{-1.6cm}
\hspace*{-1cm}
\epsfig{file=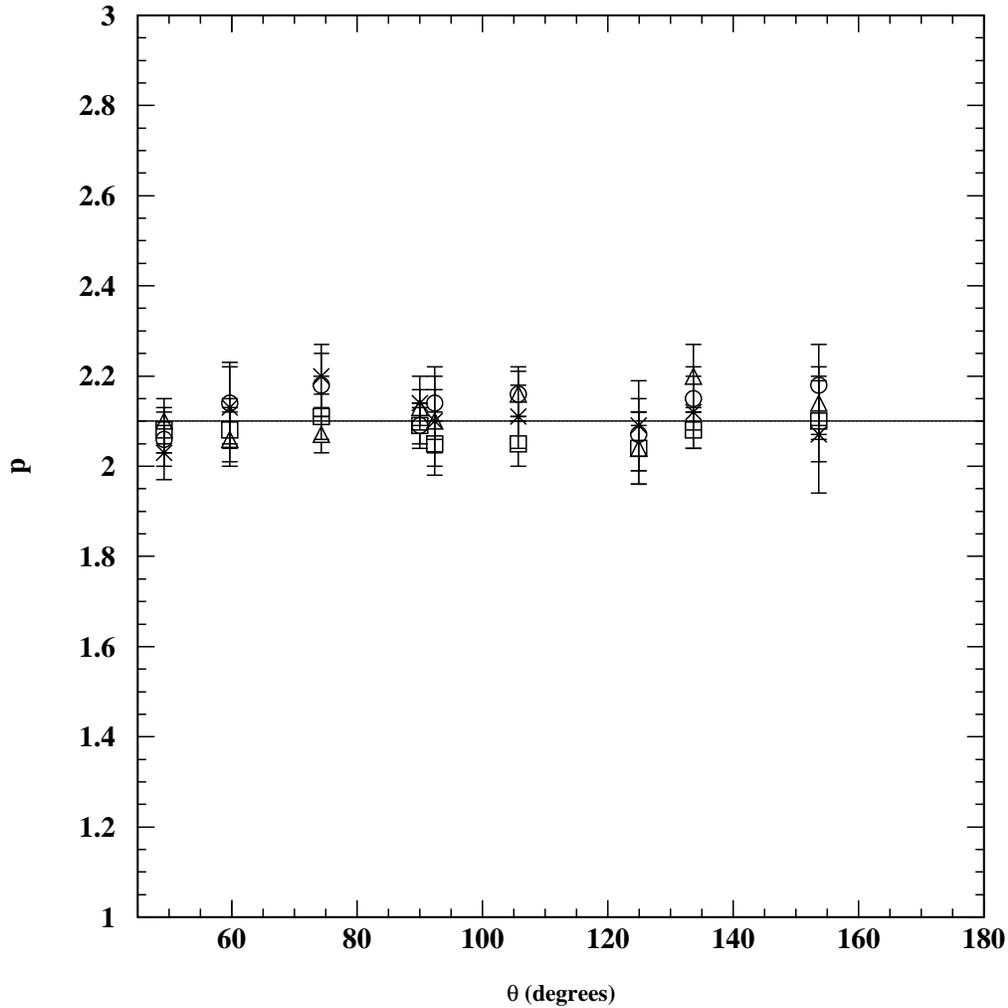,width=15cm}
\caption{EGRET data for the angular dependence of the GBR spectral 
index  as a function of direction, plotted as a function of $\theta$:
the angle away from the direction of the galactic center. 
The galactic disk and galactic bulge are masked: the data are for
${\rm |b|\geq 10^0}$ for $ {\rm |l|\geq 40^0}$ and
$ {\rm |b|\geq 30^0}$ for $ {\rm |l|\leq 40^0}$. The four points at each
$\theta$ correspond to the four half-hemispheres.
The straight line is the predicted spectral index.}
\vspace*{-0.5cm}
\end{center}
\end{figure}

\newpage
\begin{figure}
\begin{center}
\vspace*{-1.6cm}
\hspace*{-1cm}
\epsfig{file=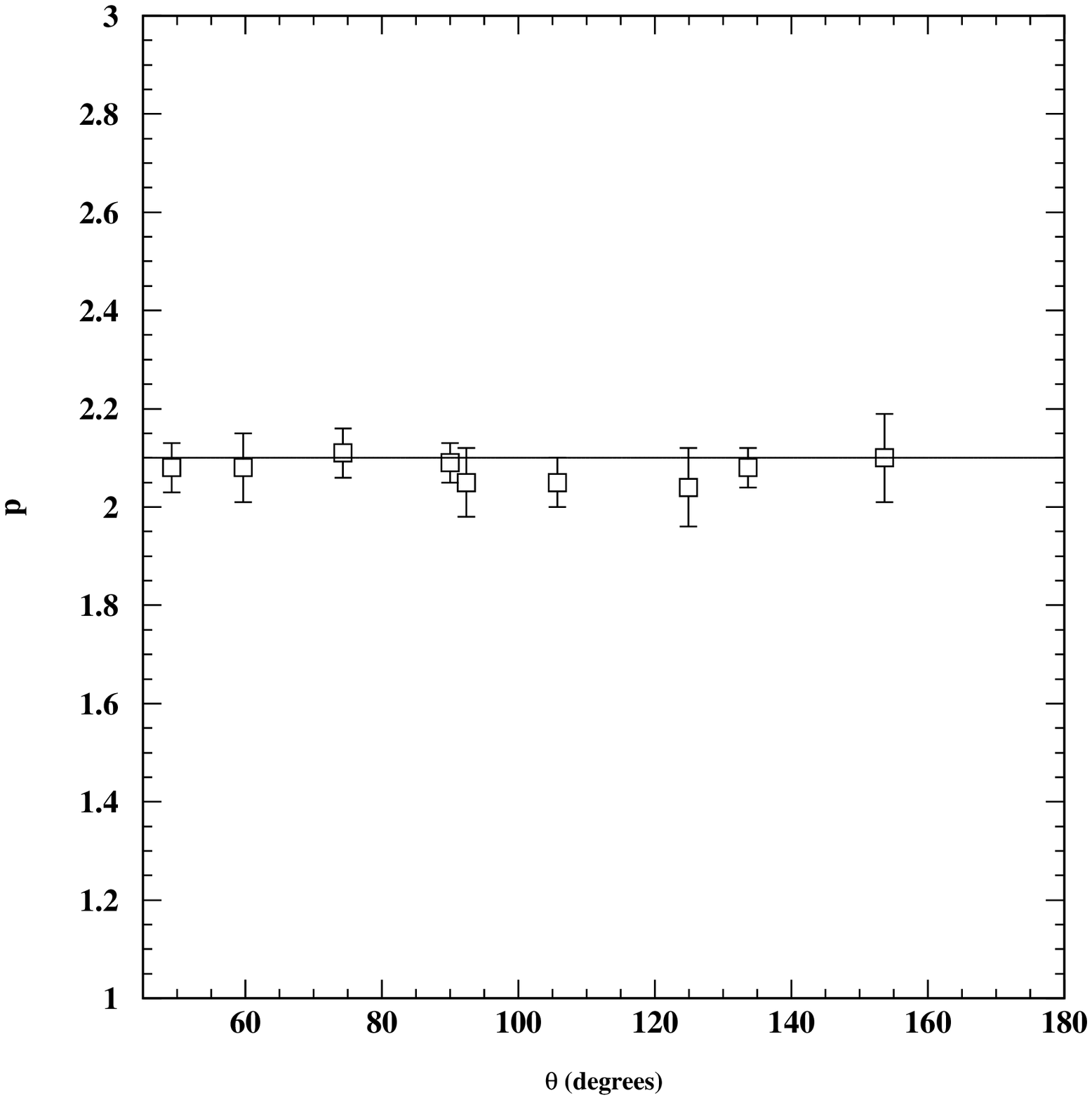,width=15cm}
\caption{Same as Fig.~3 for the half hemisphere with
$\rm{ b> 0}$, ${\rm l> 0}$.}
\vspace*{-0.5cm}
\end{center}
\end{figure}

\newpage
\begin{figure}
\begin{center}
\vspace*{-1.6cm}
\hspace*{-1cm}
\epsfig{file=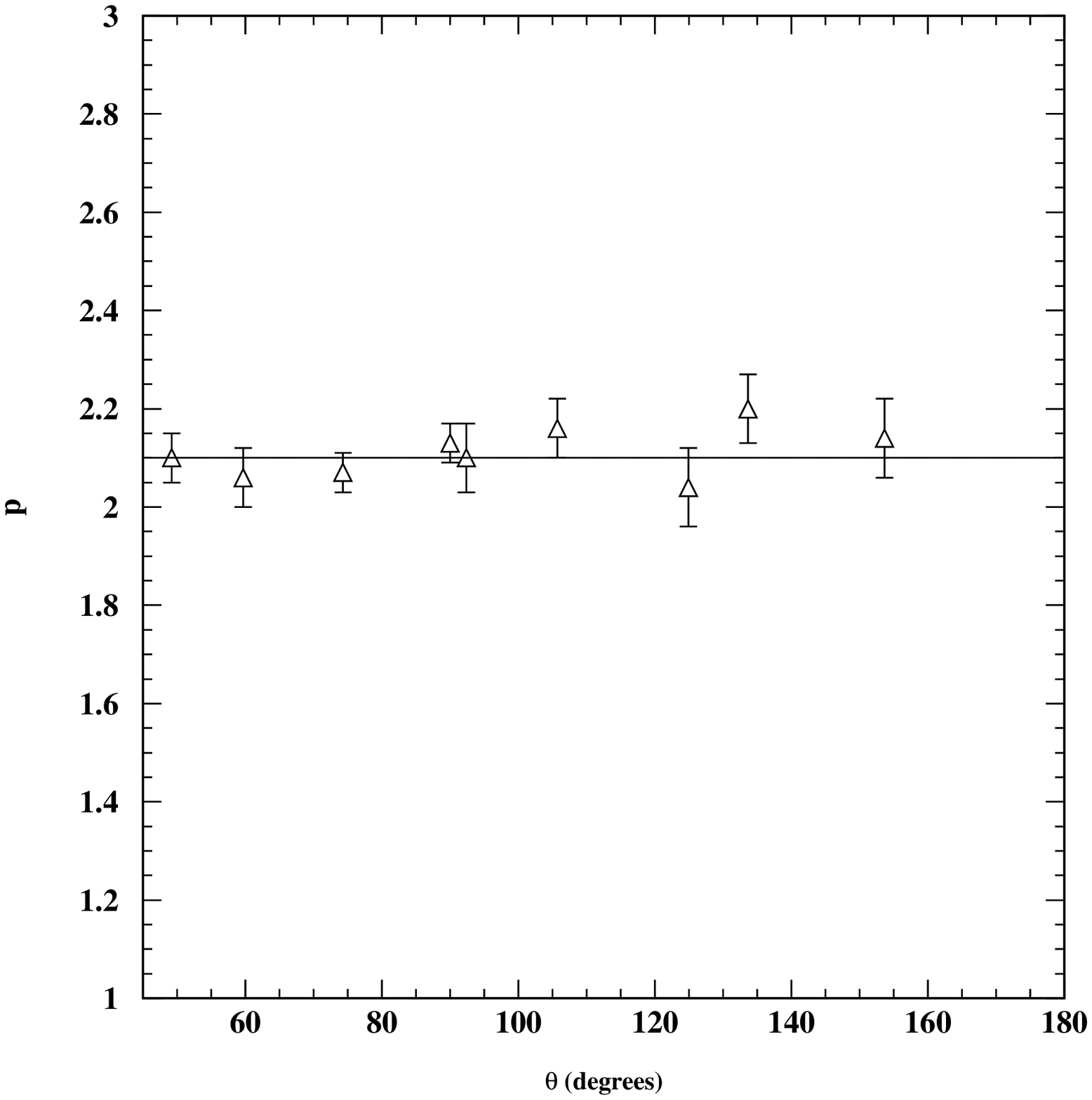,width=15cm}
\caption{Same as Fig.~3 for the half hemisphere with
${\rm b> 0}$, ${\rm l< 0}$.}
\vspace*{-0.5cm}
\end{center}
\end{figure}

\newpage
\begin{figure}
\begin{center}
\vspace*{-1.6cm}
\hspace*{-1cm}
\epsfig{file=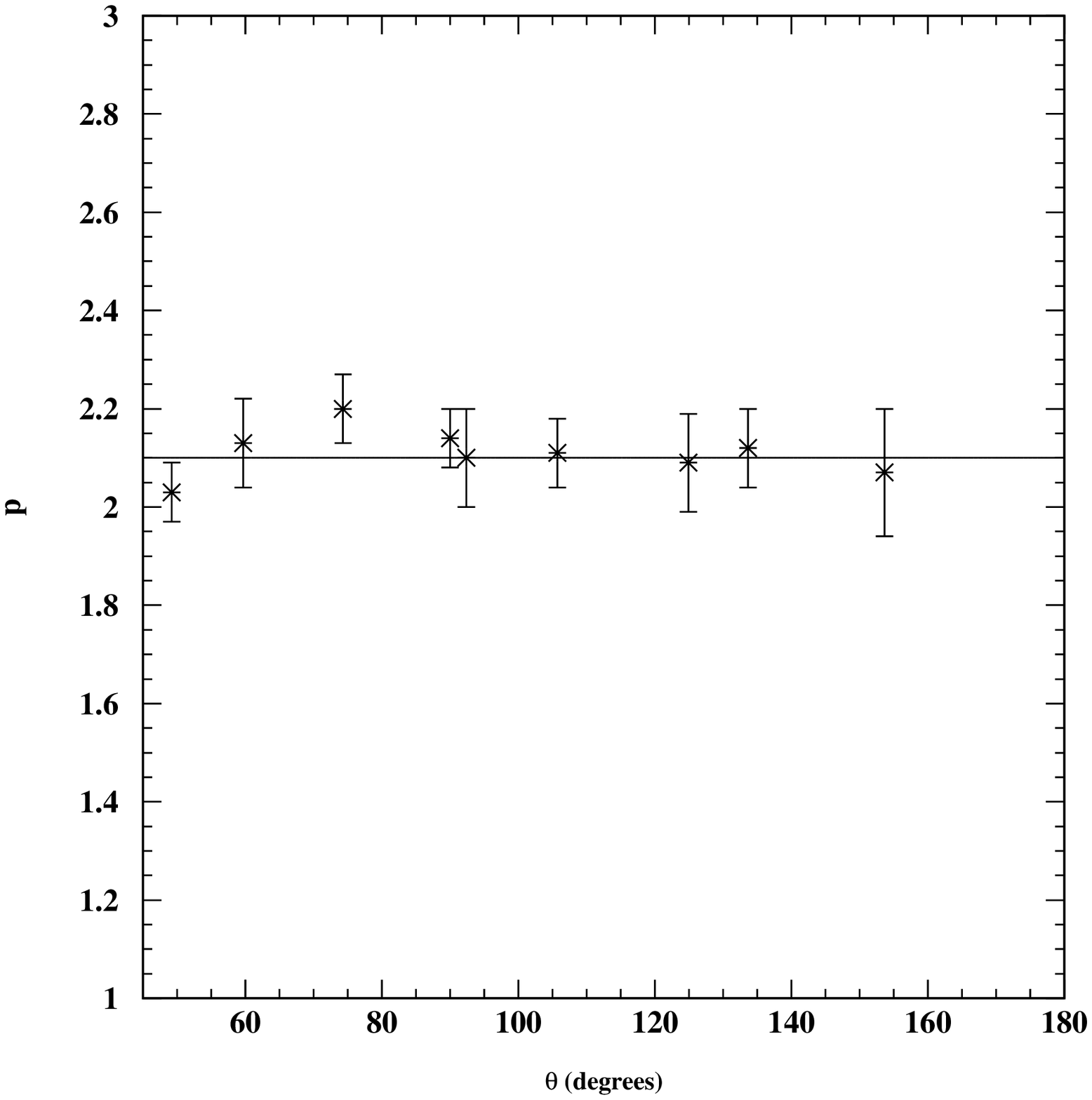,width=15cm}
\caption{Same as Fig.~3 for the half hemisphere with
${\rm b< 0}$, ${\rm l> 0}$.}
\vspace*{-0.5cm}
\end{center}
\end{figure}

\newpage
\begin{figure}
\begin{center}
\vspace*{-1.6cm}
\hspace*{-1cm}
\epsfig{file=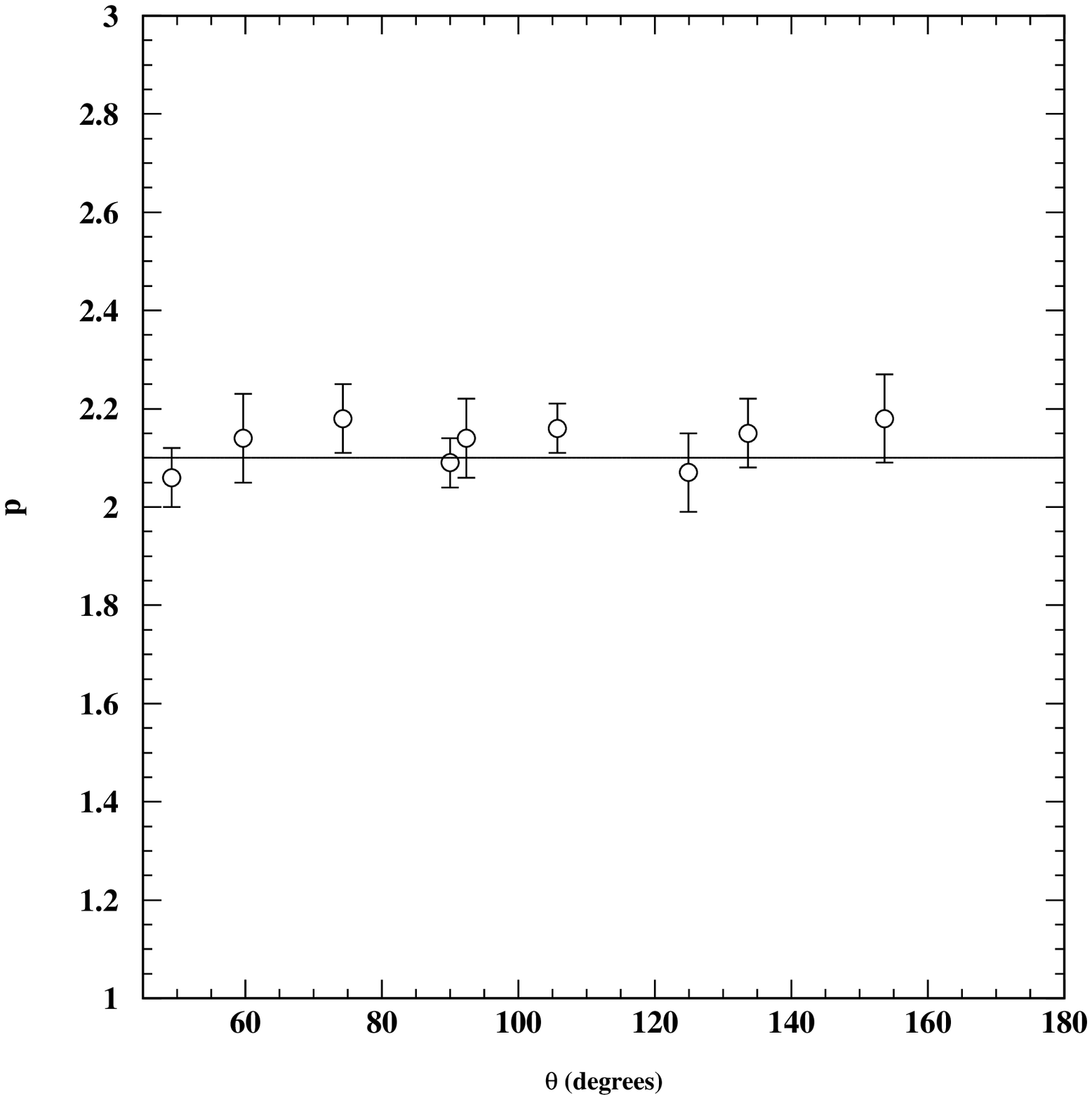,width=15cm}
\caption{Same as Fig.~3 for the half hemisphere with
${\rm b< 0}$, ${\rm l< 0}$.}
\vspace*{-0.5cm}
\end{center}
\end{figure}

\newpage
\begin{figure}
\begin{center}
\vspace*{-1.6cm}
\hspace*{-1cm}
\epsfig{file=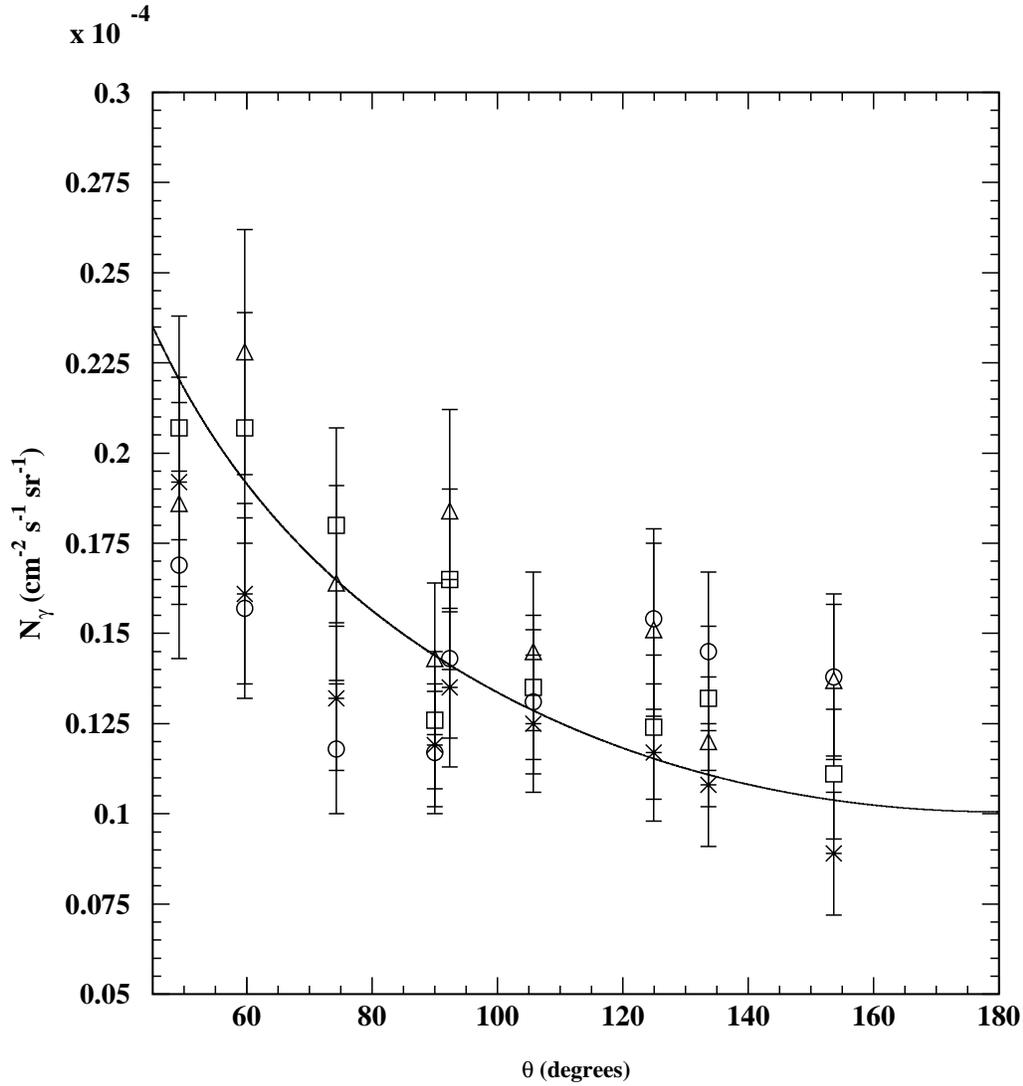,width=15cm}
\caption{EGRET data, masked and organized as in Fig.~3, 
for the dependence of the GBR intensity on $\theta$. The line is the   
prediction for ICS of starlight and the CMB by CR
electrons, for ${\rm R_e}=30$ kpc.}
\vspace*{-0.5cm}
\end{center}
\end{figure}

\newpage
\begin{figure}
\begin{center}
\vspace*{-1.6cm}
\hspace*{-1cm}
\epsfig{file=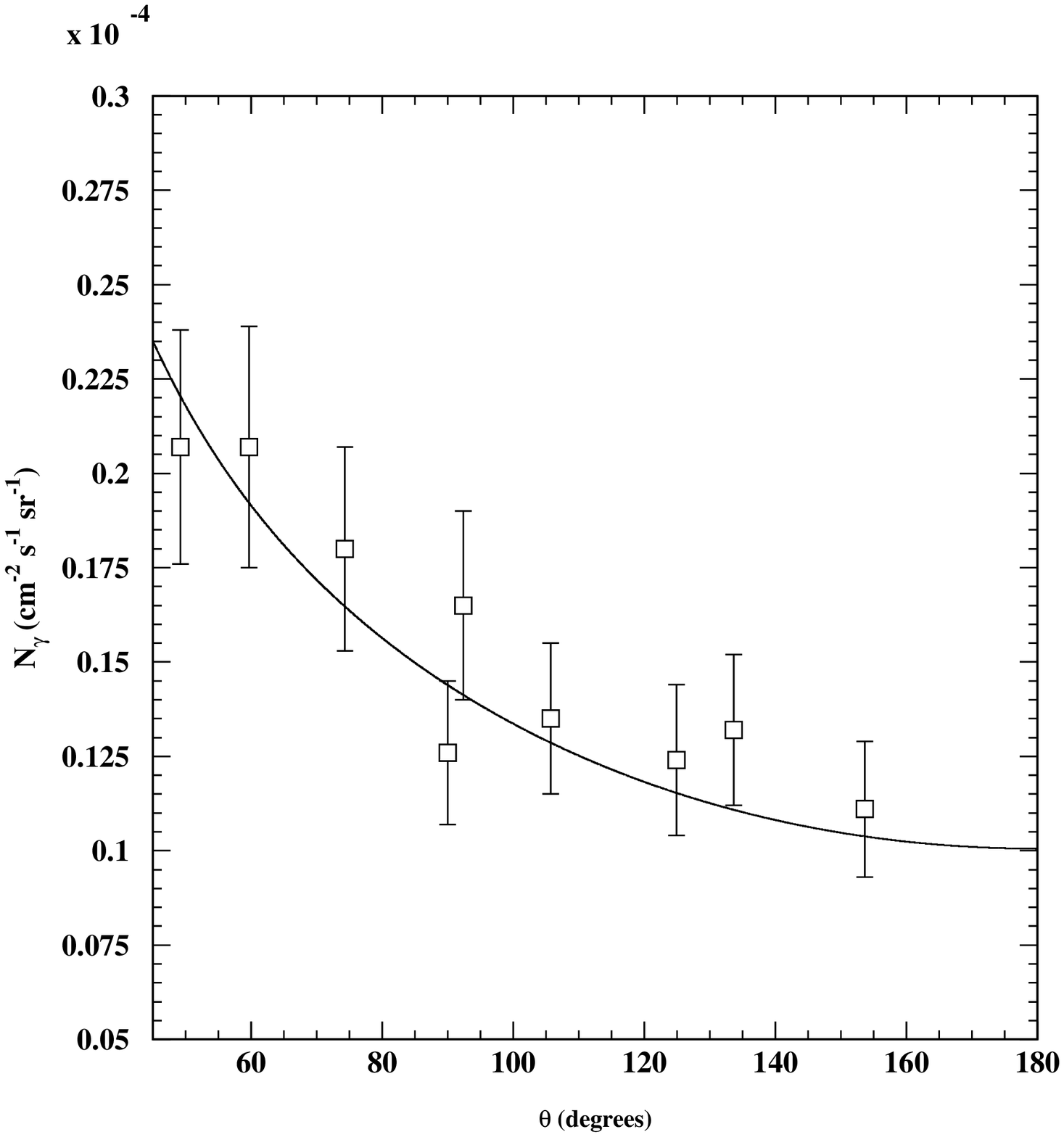,width=15cm}
\caption{Same as Fig.~8 for the half hemisphere with
$\rm{ b> 0}$, ${\rm l> 0}$.}
\vspace*{-0.5cm}
\end{center}
\end{figure}

\newpage
\begin{figure}
\begin{center}
\vspace*{-1.6cm}
\hspace*{-1cm}
\epsfig{file=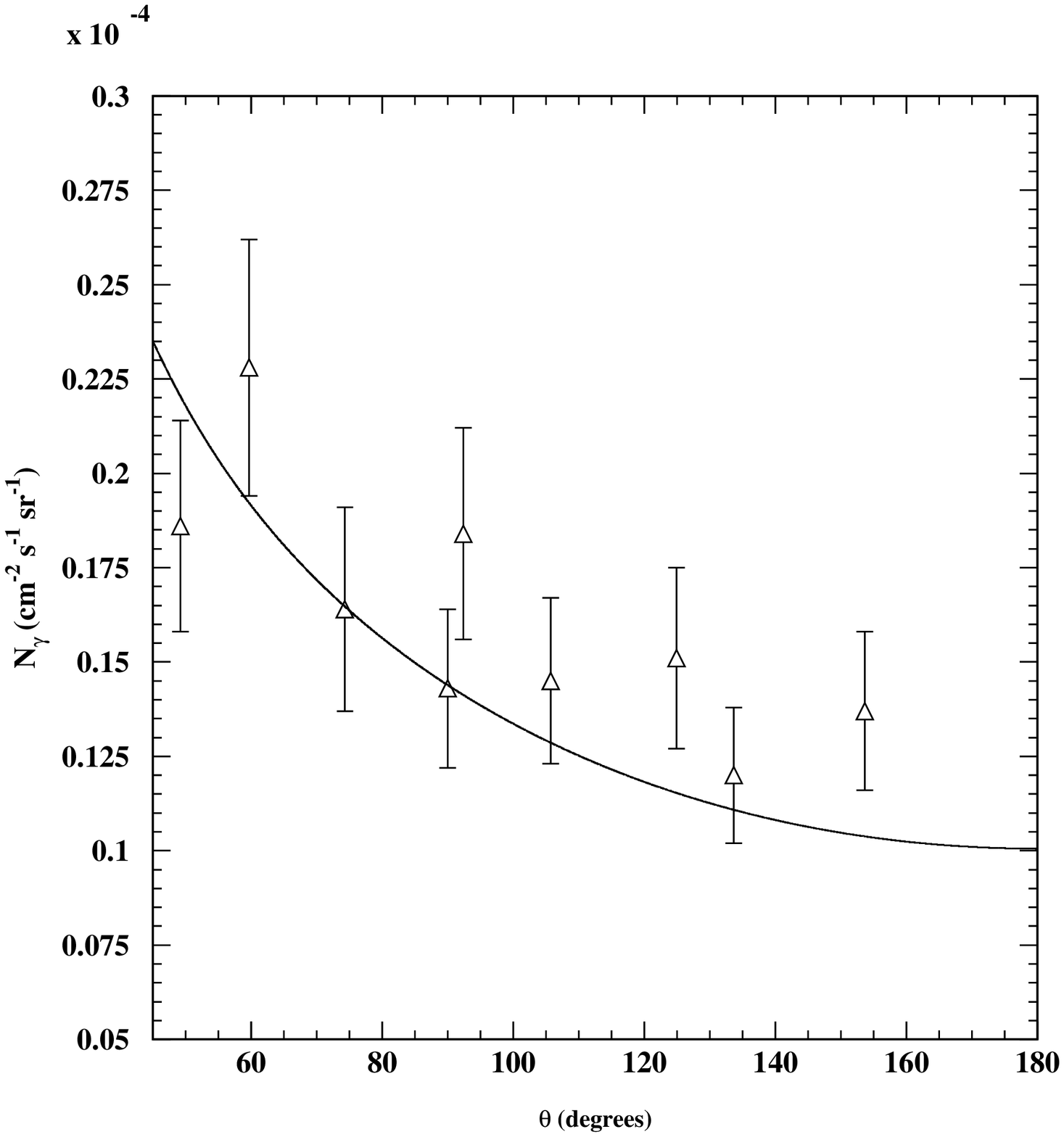,width=15cm}
\caption{Same as Fig.~3 for the half hemisphere with
${\rm b> 0}$, ${\rm l< 0}$.}
\vspace*{-0.5cm}
\end{center}
\end{figure}

\newpage
\begin{figure}
\begin{center}
\vspace*{-1.6cm}
\hspace*{-1cm}
\epsfig{file=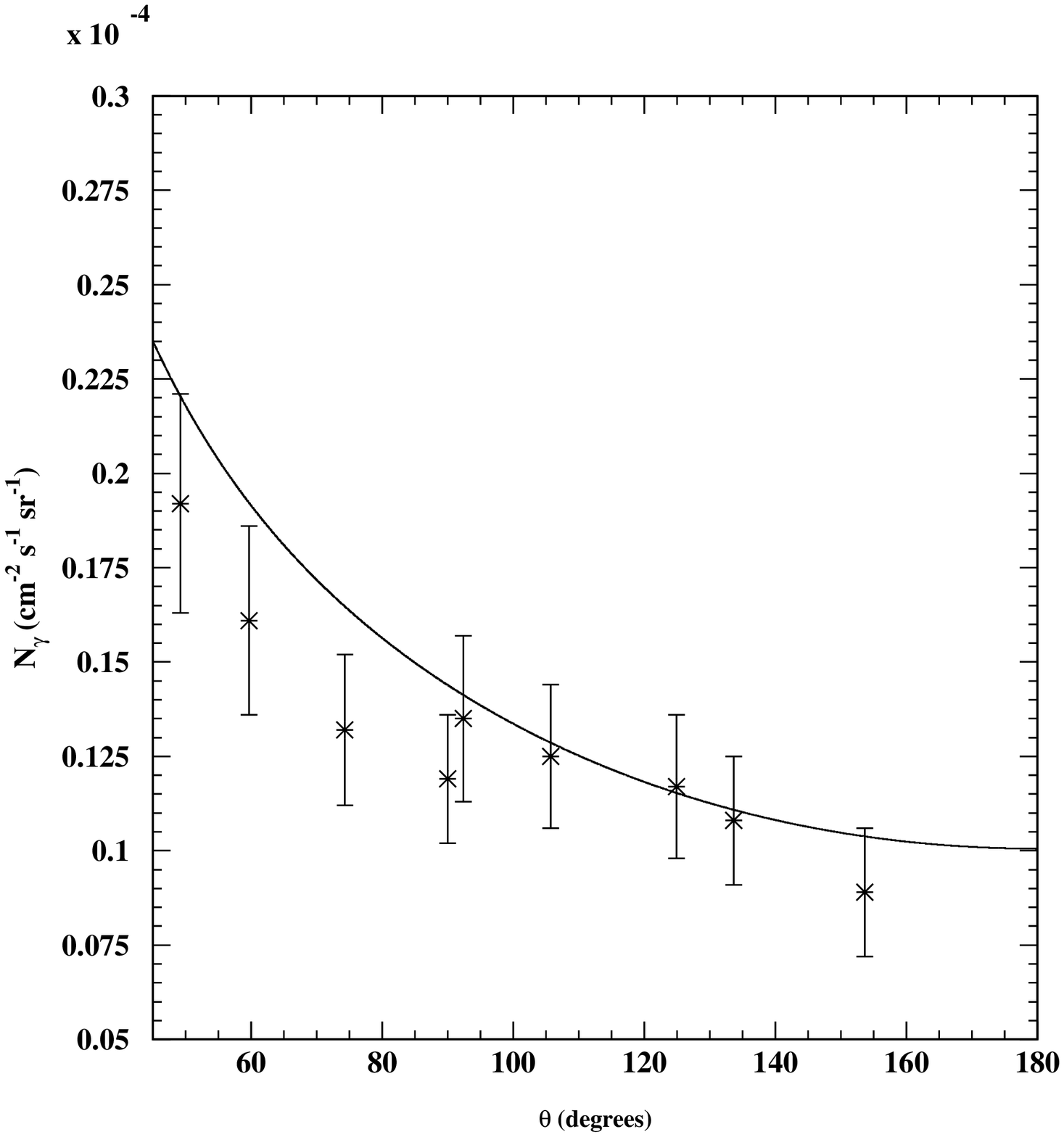,width=15cm}
\caption{Same as Fig.~8 for the half hemisphere with
${\rm b< 0}$, ${\rm l> 0}$.}
\vspace*{-0.5cm}
\end{center}
\end{figure}

\newpage
\begin{figure}
\begin{center}
\vspace*{-1.6cm}
\hspace*{-1cm}
\epsfig{file=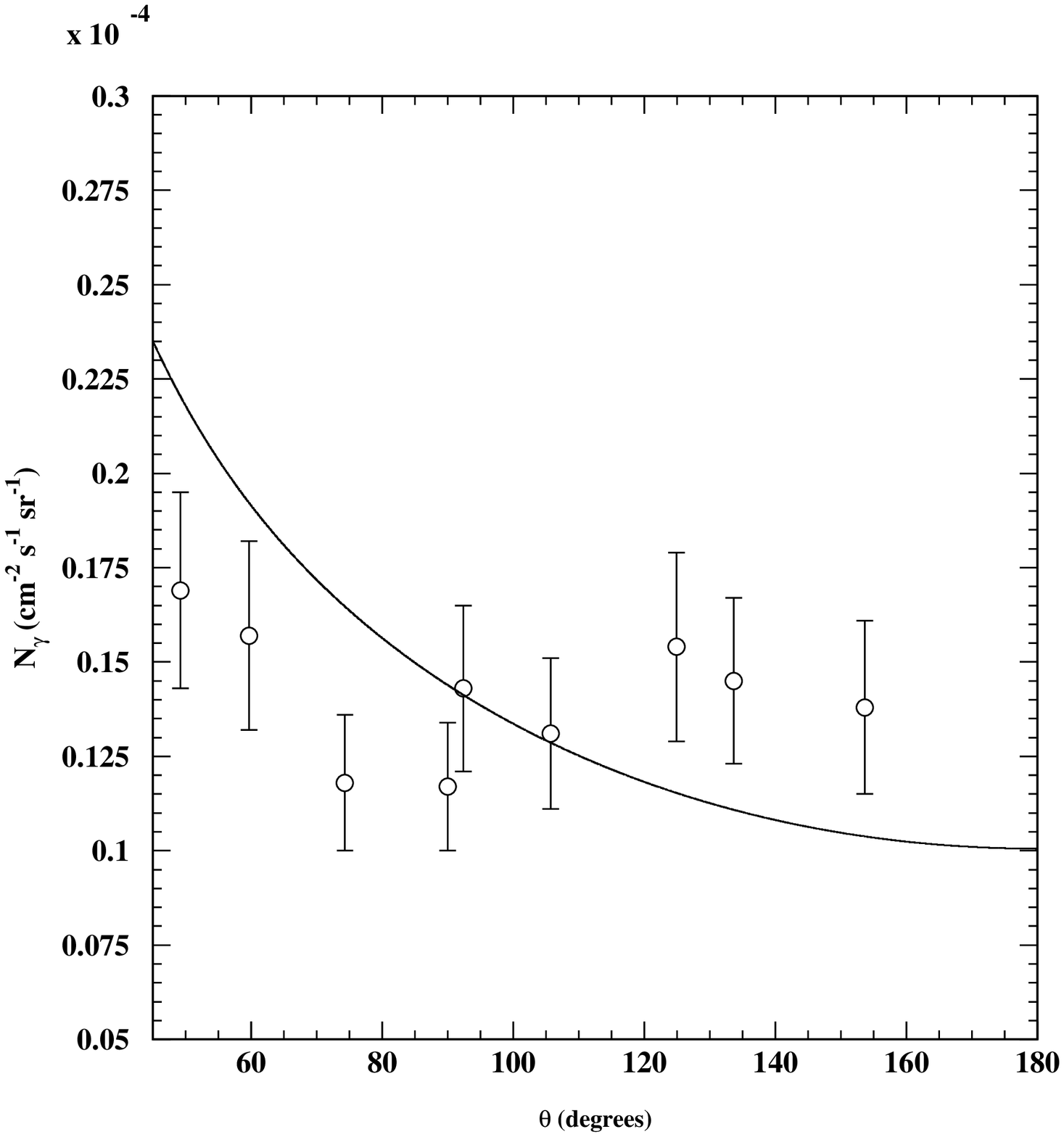,width=15cm}
\caption{Same as Fig.~8 for the half hemisphere with
${\rm b< 0}$, ${\rm l< 0}$.}
\vspace*{-0.5cm}
\end{center}
\end{figure}


\begin{thebibliography}{99}

\bibitem{BG} See, e.g., V.S. Berezinskii {\it et al.,}  {\it Astrophysics
of cosmic rays} (North Holland, Amsterdam, 1990).

\bibitem{Protheroe} For a recent review, see, e.g., P. Bhattacharjee
and G. Sigl, submitted to Phys. Rep. (astro-ph/9811011).

\bibitem{hecr} D.J. Bird et al., Astroph. J. {\bf 441}, 144 (1995)
and references therein. M. Takeda {\it et al.,} Phys. Rev. Lett. {\bf 81}, 
1163 (1998).


\bibitem{H?} F. Halzen {\it et al.,} Astropart. Phys.
{\bf 3}, 151 (1995).
J.W. Elbert and P. Sommers, Astroph. J. {\bf 441}, 151 (1995).

\bibitem{Lee} S. Lee, Phys. Rev. {\bf D58}, 043004 (1997).


\bibitem{GZK} K. Greisen, Phys. Rev. Lett. {\bf 16}, 748 (1966).
G.T. Zatsepin and V.A. Kuz'min, JETP Lett. {\bf 4}, 78 (1966).

\bibitem{eflux0} For a recent compilation see, e.g.,
B. Weibel-Sooth and P.L. Biermann,
{\it Landolt-Bornstein} (Springer Verlag, Heidelberg 1998, in press).

\bibitem{eflux1}
T.A. Prince, Astroph. J. {\bf 227}, 676 (1979).
J. Nishimura {\it et al.,} Astroph. J. {\bf 238}, 394 (1980).
K.K. Tang, Astroph. J. {\bf 278}, 881 (1984).
R.L. Golden {\it et al.,} Astroph. J. {\bf 287}, 622 (1984).
P. Evenson and P. Meyer, J. Geophys. Res. {\bf 89 A5}, 2647 (1984).
R.L. Golden {\it et al.,} Astroph. J. {\bf 436}, 739 (1994).
P. Ferrando {\it et al.,} Astron. and Astroph. {\bf 316}, 528 (1996).  
S.W. Barwick {\it et al.,} Astroph. J. {\bf 498}, 779 (1998).

\bibitem{Sree} P. Sreekumar {\it et al.,} Astroph. J. {\bf 494}, 523
(1998) and references therein.

\bibitem{Broad} A. Broadbend, C.G.T. Haslam and J.L. Osborne,
MNRAS {\bf 237}, 381 (1989)

\bibitem{Duric} N. Duric, J. Irwin and H. Bloemen,
 Astron. Astroph. {\bf 331}, 428 (1998), and references therein.

\bibitem{DP} A. Dar and R. Plaga, to appear in  Astron. Astroph.
In this paper a CR population permeating a magnetized region
of galactic-halo size is proposed as a solution to the GZK conundrum, 
as a qualitative description of the nuclear CR spectrum, and as a
link between cosmic rays and gamma-ray bursts.

\bibitem{Simp} See, e.g., J.A. Simpson and J.J. Connel,
 Astroph. J. {\bf 497}, L88 (1998).

\bibitem{BO}  J. Bednarz and M. Ostrowski,
Phys. Rev. Lett. {\bf 80}, 3911 (1998).

\bibitem{D98} A. Dar, astro-ph/9809163, in Proceedings of the
Rencontres de la Vall\'ee d'Aoste, 1998 (ed. M. Greco),
Frascati Physics Series, INFN Pubs, page 23.


\bibitem{Swordy} See, for instance,
S.P. Swordy {\it et al.,} Astroph. J. {\bf 330}, 625 (1990).

\bibitem{Barb}
R.L. Golden {\it et al.,} Astroph. J. {\bf 436}, 739 (1994).
G. Barbiellini {\it et al.,}  Astron. and Astroph. {\bf 309}, L15 (1996).
S.W. Barwick {\it et al.,} Astroph. J. {\bf 498}, 779 (1998).
S. Coutu {\it et al.}, to appear in Astropar. Phys.
(astro-ph/99002162).

\bibitem{MF} J.C. Mather {\it et al.,}  Astroph. J. {\bf 432}, L15 (1993).
D. Fixsen {\it et al.,} Astroph. J. {\bf 473}, 576 (1996) and references 
therein.

\bibitem{TF} D.J. Thompson and C.E. Fichtel, Astron. and Astroph.
{\bf 109}, 352 (1982).


\bibitem{PE98} M. Pohl and J.A. Esposito, Astroph. J. {\bf 507}, 327  
(1998).


\bibitem{Strong} A.W. Strong {\it et al.,} Proc. 4th Compton Symp. AIP, 
{\bf 410}, 1198 (1997).

\bibitem{Moska} See, for instance,
I.V. Moskalenko and A.W. Strong, astro-ph/9811221 and
references therein.


\bibitem{FMRL}  J.E.  Felten  and P. Morrison,
Astroph. J. {\bf 146}, 686 (1966).

\bibitem{Hun} S.D. Hunter {\it et al.,} Astroph. J. {\bf 481}, 205 (1997).

\bibitem{Love} J. Loveday {\it et al.,} Astroph. J. {\bf 390}, 338 (1992).

\bibitem{Wijers} See, for instance,
R.A.M.J. Wijers {\it et al.,} MNRAS {\bf 294}, L13 (1997).


\bibitem{S} C.C. Steidel {\it et al.,} astro-ph/9811399.

\bibitem{SS} M.H. Salamon and F.W. Stecker,
 Astroph. J. {\bf 493}, 547 (1998).






\end{thebibliography}
\end{document}